\documentclass[prl,aps,showpacs,twocolumn,unsortedaddress]{revtex4}
\usepackage{graphics,bm}
\usepackage{amssymb}
\usepackage{epsfig}
\usepackage{epsf}

\begin{document}

\title{Antiferromagnetic Metal Spintronics}

\author{A. S. N\'u\~nez}
\email{alnunez@physics.utexas.edu}
\homepage{http://www.ph.utexas.edu/~alnunez}

\author{R.A. Duine}
\email{duine@physics.utexas.edu}
\homepage{http://www.ph.utexas.edu/~duine}

\author{A.H. MacDonald}
\email{macd@physics.utexas.edu}
\homepage{http://www.ph.utexas.edu/~macdgrp}

\affiliation{The University of Texas at Austin, Department of
Physics, 1 University Station C1600, Austin, TX 78712-0264}

\date{\today}

\begin{abstract}
Spintronics in ferromagnetic metals is built on a complementary
set of phenomena in which magnetic configurations influence
transport coefficients and transport currents alter magnetic
configurations. In this Letter we propose that corresponding
effects occur in circuits containing antiferromagnetic metals. The
critical current for switching can be smaller in the
antiferromagnetic case because of the absence of shape anisotropy
and because spin torques act through the entire volume of an
antiferromagnet. Our findings suggest that current-induced order
parameter dynamics can be used to coarsen the microstructure of
antiferromagnetic thin films.
\end{abstract}
\vskip2pc

\maketitle

\noindent {\em Introduction} --- Spintronics in ferromagnetic
metals\cite{Review} is based on one hand on the dependence of
resistance on magnetic microstructure \cite{GMRrefs}, and on the
other hand on the ability to alter magnetic microstructures with
transport currents
\cite{Slon,Berger,Tsoi1,Tsoi2,Sun,SMT-exp,Chien,MSU}.  These
effects are often largest and most robust in circuits containing
ferromagnetic nanoparticles that have a spatial extent smaller
than a domain wall width and therefore largely coherent
magnetization dynamics. In this Letter we point out that similar
effects occur in circuits containing antiferromagnetic metals. The
systems that we have in mind are antiferromagnetic transition
metals similar to Cr\cite{FawcettI} and its alloys\cite{FawcettII}
or the rock salt structure intermetallics \cite{ExchangeBias} used
as exchange bias materials which are well described by
time-dependent mean-field-theory in its density-functional
theory\cite{Gunnarrson} setting.

Our proposal that currents can alter the micromagnetic state of an
antiferromagnet may seem surprising since spin-torque effects in
ferromagnets \cite{TransferTheory} are usually discussed in terms
of conservation of total spin, a quantity that is not related to
the staggered moment order parameter of an antiferromagnet. Our
arguments are based on a microscopic picture of
spin-torques\cite{nunez2004} in which they are viewed as a
consequence of changes in the exchange-correlation effective
magnetic fields experienced by all quasiparticles in the transport
steady state. A spin torque that drives the staggered-moment
orientation $\mathbf{n}$ must also be staggered, and will be produced\cite{nunez2004}
by the exchange potential due to an {\em unstaggered}
transport electron spin-density in the plane perpendicular to
$\mathbf{n}$.  The required alteration in torque is produced by the
alternating moment orientations in the antiferromagnet
rather than the transport electron exchange
field.  As we now explain the transverse
spin-densities necessary for a staggered torque
occur generically in circuits containing
antiferromagnetic elements.

\begin{figure}
\vspace{-0.5cm}
\centerline{\epsfig{figure=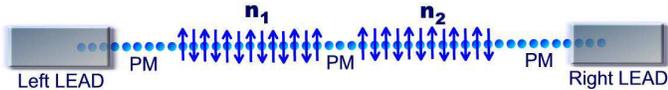,width=9.0cm}}
 \caption{ The model heterostructure for which we perform our
 calculations.
  } \label{fig:cartoon}
\end{figure}

The key observations behind our theory
concern the scattering properties of a single channel containing
non-collinear antiferromagnetic elements with a staggered exchange
field that varies periodically along the channel and is
commensurate with an underlying lattice that has inversion
symmetry.  For an antiferromagnetic element that is invariant
under simultaneous spatial and staggered moment inversion it
follows from standard one-dimensional scattering theory
\cite{Melo} considerations that transmission through an individual
antiferromagnetic element is spin-independent, and that the
spin-dependent reflection amplitude from the antiferromagnet or
any period thereof has the form ${\bf r} = r_s \mathbf{1} + r_t \;
{\bf n} \cdot \vec{\tau}$, where ${\bf n}$ is the order parameter orientation
and $\vec{\tau}$ are the
Pauli spin matrices; $r_s$ and $r_t$ are proportional to sums and
differences of reflection amplitudes for incident spins oriented
along and opposite to the staggered moment.  The reflection
amplitude for a spinors incident from opposite sides differ by changing
the sign of ${\bf n}$ and the transmission amplitudes are identical.
It then follows from composition rules for
transmission and reflection amplitudes in a compound circuit
containing paramagnetic source and drain electrodes and two
antiferromagnetic elements with staggered moment orientations
${\bf n}_1$ and ${\bf n}_2$ separated by a paramagnetic spacer
(see Fig.~\ref{fig:cartoon}) that the transport electron
spin-density in the ${\bf n}_1 \times {\bf n}_2$ direction is
periodic in the antiferromagnets. (We define the direction of ${\bf
n}_i$ to be the direction of the local moment opposite the
spacer.) The spin-torques that appear in this type of circuit
therefore {\em act through the entire volume of each
antiferromagnet}.

A proof of this property will be presented elsewhere.  Here we
illustrate the potential consequences of this property by using
non-equilibrium Greens function techniques to evaluate
antiferromagnetic giant magnetoresistance (AGMR) effects and
layer-dependent spin-torques in model two-dimensional circuits
containing paramagnetic and antiferromagnetic elements.  We focus
on the most favorable case in which the antiferromagnet has a
single ${\bf Q}$ spin-density-wave state with ${\bf Q}$ in the
current direction.  In the following we first explain the model
system that we study and the non-equilibrium Greens function
calculation that we use to evaluate magnetoresistance and
spin-torque effects.  We conclude that under favorable
circumstances, both effects can be as large as the ones that occur
in ferromagnets. We then estimate typical critical current for
switching an antiferromagnet. Finally, we discuss some of the
challenges that stand in the way of realizing these effects
experimentally.

\noindent {\em Antiferromagnetic giant magnetoresistance} --- We
start by analyzing the simplified two-dimensional lattice model of
an antiferromagnetic heterostructure characterized by
near-neighbor hopping, transverse translational invariance, and
spin-dependent on-site energies, that is illustrated in
Fig.~\ref{fig:cartoon}:
\begin{eqnarray}\label{eq:singlepartHamiltonian}
\mathcal{H}_k &=&  - t \sum_{\langle i,j \rangle,\sigma}
c^\dagger_{k,i,\sigma}\;
c^{\phantom{\dagger}}_{k,j,\sigma}+ \mbox{h.c.} \nonumber \\
&+&\sum_{i,\sigma,\sigma^\prime} \left[ ( \epsilon_i+\epsilon_k )
\delta_{\sigma,\sigma^\prime}- \Delta_i \hat
\mathbf{{\Omega}}_i\cdot\vec{\tau}_{\sigma,\sigma^\prime} \right]
c^\dagger_{k,i,\sigma}\; c^{\phantom{\dagger}}_{k,i,\sigma'}~.
\end{eqnarray}
Here, $k$ denotes the transverse wave number, $t$ the hopping
amplitude and $\epsilon_k$ the transverse kinetic energy. The second term in
Eq.~(\ref{eq:singlepartHamiltonian}) describes the exchange
coupling $\Delta_i$ of electrons to antiferromagnetically ordered
local moments $\hat {\mathbf \Omega}_i = (-)^{i} \mathbf{n}$ that alternate
in each antiferromagnet. In the paramagnetic
regions of these model systems $\Delta_i = 0$.  The on-site
energies $\epsilon_i$ are allowed to change across a
heterojunction.

\begin{figure}
\centerline{\epsfig{figure=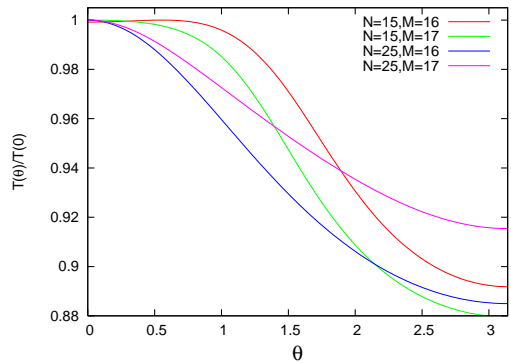,width=7.0cm}}
\caption{Landauer-Buttiker conductance as a function of the angle
$\theta$ between the magnetization orientations $\hat {\mathbf
\Omega}_i$ on opposite sides of the paramagnetic spacer layer.
There is a sizable giant magnetoresistance effect, with larger
conductance at smaller $\theta$ and weak dependence on layer
thicknesses. These results were obtained for $\Delta/t=1$ and
$\epsilon_i=0$. } \label{fig:transmission}
\end{figure}

We use the non-equilibrium Greens function formalism to describe
the transport of quasiparticles across the magnetic
heterostructure. The essential physical properties of the system
are encoded in the real time Greens function
\cite{caroli1972,dattabook}, defined by the ensemble  average,
$G^{<}_{\sigma,i;\sigma^\prime,j} (k;t,t^\prime)= {\mathbf i}
\langle c^\dagger_{k,i,\sigma} (t)\;
c^{\phantom{\dagger}}_{k,j,\sigma}(t^\prime)\rangle$, from which
the (spin) current and (spin) density can be evaluated. To
determine the model's AGMR effect, we calculate the transmission
coefficient as a function of the angle $\theta$ between
orientations $\hat {\mathbf \Omega}_i$ on opposite sides of the
spacer.  In Fig.~\ref{fig:transmission} the transmission
coefficient is shown for specific values of the number of layers
$N$ and $M$, in the first and second antiferromagnet. The AGMR
effect can be traced to the interference between spin-current
carrying electron spinors reflected by the facing layers. (This is
also the origin of spin transfer.)  For the model we study the AGMR
depends on the orientation of the layers opposite the spacer in
the usual way, {\em i.e.} the resistance is highest for
$\theta=\pi$ and lowest for $\theta=0$. Also, we find that the
AGMR ratio, defined as the absolute difference between the maximum
and minimum value of the transmission coefficient normalized to
the minimum, saturates as a function of the length of the
antiferromagnets.

\begin{figure}
\centerline{\epsfig{figure=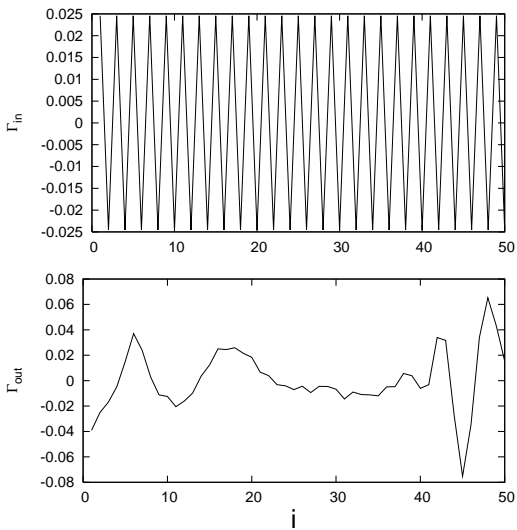, width=7cm}} \caption{Local
spin-transfer torques in the down-stream antiferromagnet.  The
in-plane spin transfer is staggered and therefore effective in
driving coherent order parameter dynamics. The out-of-plane
spin-transfer is locally up to one order of magnitude larger, but
is ineffective because it is not staggered.  These results were
obtained for $\Delta/t=1$, $\epsilon_i=0$, $\theta=\pi/2$, $N=50$,
and $M=50$.} \label{fig:localST}
\end{figure}

\noindent {\em Current-driven switching of an antiferromagnet} ---
To address the possibility of current-induced switching of an
antiferromagnet we evaluate spin transfer torques in the second
antiferromagnet. The spin transfer torque originates from the
contribution made by transport electrons to the
exchange-correlation effective magnetic field and is
given\cite{nunez2004} by ${\mathbf \Gamma}=\Delta_i \hat {\mathbf
\Omega}_i \times \langle {\bf s}_i\rangle/\hbar$, where $\langle
{\bf s}_i\rangle$ is the nonequilibrium expectation value of the
quasiparticle spin.  We distinguish the spin-torque component in
the plane spanned by ${\bf n}_1$ and ${\bf n}_2$ and the component
out of this plane. In Fig.~\ref{fig:localST} we show the in-plane
and out-of-plane transport-induced spin torques. As anticipated the
in-plane spin transfer torque in this model is {\em exactly}
staggered and is therefore extremely effective in driving
order-parameter dynamics. We have checked numerically that
staggered in-plane spin-transfer torques that do not decay also
occur in continuum toy models of an antiferromagnet with
piece-wise constant and sinusoidal exchange fields.  These
persistent spin torques are a generic property of
antiferromagnetic circuits related to the absence of
spin-splitting in the Bloch bands. The
staggered in-plane spin-transfer is produced by an out-of-plane
spin density that is {\em exactly} constant in our lattice model
antiferromagnet and exactly periodic in a continuum model
antiferromagnet.

\begin{figure}
\vspace{-0.5cm} \centerline{\epsfig{figure=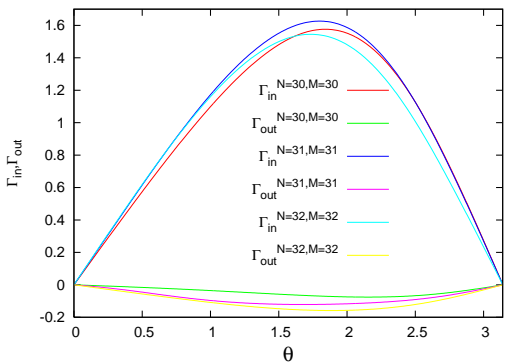,width=7.0cm}}
 \caption{Total spin transfer torque action on the downstream antiferromagnet,
 as a function of $\theta$. We used the parameters $\Delta/t=1$ and $\epsilon_i=0$.} \label{fig:STvsQ}
\end{figure}

\begin{figure}
\vspace{-0.5cm}
\centerline{\epsfig{figure=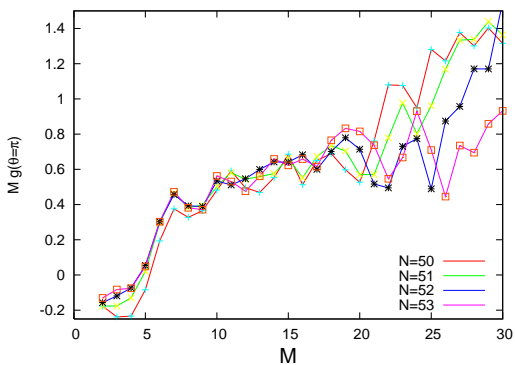,width=7.0cm}}
 \caption{Derivative of the total spin transfer torque per unit current, $M g(\theta=\pi)$, acting on the downstream antiferromagnet
  with respect to the angle $\theta$ at $\theta=\pi$ as a function of $M$. We used the parameters $\Delta/t=1$ and $\epsilon_i=0$.} \label{fig:g}
\end{figure}

If the exchange-interactions that stabilize the antiferromagnetic
state are very strong, the magnetization dynamics of each antiferromagnetic
element will be coherent and respond only to the staggered component of each spin-torque.
In Fig.~\ref{fig:STvsQ} we show the total staggered torque acting
on the downstream antiferromagnet, as a function of the angle
$\theta$. Clearly, the out-of-plane component of the torque is
small compared to the in-plane component. Fig.~\ref{fig:g} shows the derivative of the spin transfer
torque per unit current with respect to $\theta$, which we denote
$Mg(\theta)$, at $\theta=\pi$. As we will see, the critical
current for reversal is inversely proportional to this quantity.

Having demonstrated the presence of spin transfer torques in a
heterostructure containing two antiferromagnetic elements, we
estimate the critical current for switching the second
antiferromagnet assuming that the first is pinned.  To illustrate our ideas, we use the crystalline
anisotropy energy density for Cr \cite{FawcettI,fenton1978}, given
by
\begin{equation}
\label{eq:Eanis}
  E( {\bf n})= K_1 (\hat {\bf z} \cdot {\bf n})^2 + K_2  (\hat {\bf x} \cdot {\bf
  n})^2  (\hat {\bf y} \cdot {\bf
  n})^2~,
\end{equation}
where ${\bf n}$ is a unit vector in the direction of the staggered
moment. The first term changes sign at the spin flop transition
\cite{FawcettI}, and forces the staggered moment to be either
parallel or perpendicular to the ordering vector ${\bf Q}$. At
room temperature ${\bf Q} \bot {\bf n}$, and for the geometry in
Fig.~\ref{fig:cartoon} we have ${\bf Q} \parallel \hat {\bf z}$
leading to the first term in Eq.~(\ref{eq:Eanis}). The term
involving $K_2$ describes cubic anisotropy in the plane
perpendicular to ${\bf Q}$.

As we have seen, the spin transfer torques act cooperatively
throughout the entire antiferromagnet. Therefore, we can focus our
description on one ferromagnetic layer within the antiferromagnet,
since the antiferromagnet ordering will be preserved as the
antiferromagnet switches. Within this approach, the dynamics of
the staggered moment of the second antiferromagnet is analogous to
the ferromagnetic case, and its equation of motion reads
\begin{eqnarray}
\label{eq:eom}
  \frac{d {\bf n}_2}{dt} &=& {\bf n}_2 \times \left[ -\frac{\gamma}{M_s} \frac{\partial E({\bf n}_2)}{\partial {\bf n}_2}\right]
  + g (\theta) \omega_j {\bf n}_2 \times ({\bf n}_1 \times {\bf n}_2) \nonumber \\
  &&- \alpha {\bf
  n}_2 \times \frac{d{\bf n}_2}{dt}~.
\end{eqnarray}
Here, $\gamma\simeq \mu_{\rm B}/\hbar$ denotes the
gyromagnetic ratio, and $M_s \simeq \mu_{\rm B}/a^3$ denotes the
saturated staggered moment density, where $a \simeq 0.3$ nm
denotes the lattice constant of Cr. The term involving $\omega_j \equiv
\gamma \, \hbar j \, /(2 e a M_s)$, with $j$ the current density and $e$
the electron charge, describes the in-plane spin transfer torque.
We neglect the out-of-plane component because, as we have seen, it
averages to a small value. Moreover, the out-of-plane
component of the spin torque competes with the anisotropy, whereas
the in-plane component competes with the damping term.  For this reason
it turns out that, even in ferromagnets, the in-plane component of
the spin torque is most important in determining the critical current for
current-driven switching.  The last term in Eq.~(\ref{eq:eom}) describes the usual Gilbert
damping, with a dimensionless damping constant for which we take
the typical value $\alpha=0.1$ \cite{fenton1978}. The anisotropy
constants are given by $K_1=10^3$ J m$^{-3}$ and $K_2=10$ J
m$^{-3}$ \cite{fenton1978}.

A linear stability analysis of Eq.~(\ref{eq:eom}) shows that for
the optimal situation ${\bf n}_1 = -\hat {\bf x}$, the fixed point
${\bf n}_2 = \hat {\bf x}$ becomes unstable if
\begin{equation}
\label{eq:jc} j \equiv j_c = \frac{e \alpha a}{g(\pi) \hbar}
\left( K_1+8  K_2\right) \simeq 10^5 {\rm A~cm}^{-2},
\end{equation}
where the value for $g(\pi)$ is found to be $g(\pi)\simeq 0.05$.

This critical current is smaller than the typical
value for switching an ferromagnet primarily because
the spin transfer torques act cooperatively throughout the
entire antiferromagnet and also because of the absence of shape anisotropy.
Using the model of Eq.~(\ref{eq:eom}) we also find that depending on the
applied current, the staggered moment ${\bf n}_2$ can relax to stable
fixed points at ${\bf n}_2 = \pm \hat {\bf y}$ or completely
reverse its direction.

\noindent {\em Discussion and conclusions} --- The calculations we
have performed are in the ballistic regime, and we expect the AGMR
and spin transfer torque effect to occur only in sufficiently
clean samples. Since both effects rely on interferences,
however, we do not expect that disorder will make it impossible to
realize the effect in typical nanoscale layered systems.  Initial experimental
explorations of this effect might be most easily interpreted in clean epitaxially
grown materials.  The complicated antiferromagnetic domain structure, known to
play a complex role in exchange biasing materials \cite{nogues1999}, might
cause AGMR and antiferromagnetic spin transfer to be smaller
than expected on the basis of our calculation.  We point out that it might
be possible to use the effects discussed here to coarsen the domain structure
of antiferromagnetic thin films.  We therefore
expect that the metallic materials used for exchange biasing are
generally a good starting point in searching for materials
displaying these antiferromagnetic spintronics phenomena.  The materials
combinations that will exhibit the effects we have in mind most strongly
depend on a large variety of considerations and can be identified by a
combination of experimental and theoretical work which follows in the
footsteps of the successful ferromagnetic metals materials research.  Finally
we remark that related effects occur in hybrid circuits containing both
antiferromagnetic and ferromagnetic elements.

In conclusion we propose that the experimental and theoretical
study of the influence of current on microstructure in circuits
containing antiferromagnetic elements will reveal interesting new
physics only partly anticipated in this Letter, and that
microstructure changes can be sensed by resistance changes. It is
a pleasure to thank Olle Heinonen, Chris Palmstrom, and Maxim Tsoi
for helpful remarks. This work was supported by the National
Science Foundation under grants DMR-0115947 and DMR-0210383, by a
grant from Seagate Corporation, and by the Welch Foundation.

\end{document}